\documentclass[prd,amsfonts,twocolumn,nofootinbib,showpacs]{revtex4}
\usepackage{graphicx}% Include figure filestau
\usepackage{dcolumn}% Align table columns on decimal point
\usepackage{bm}% bold math
\usepackage{amsmath}
\usepackage{amsfonts} 
\usepackage{amssymb}
\usepackage{color}
\usepackage{aas_macros}

\begin{document}

\title{Clusters of galaxies and variation of the fine structure constant}

\author{Silvia Galli$^{a,b}$ }
\affiliation{$^a$ UPMC Univ Paris 06, UMR7095, Institut d'Astrophysique de Paris, F-75014, Paris, France}
\affiliation{$^b$ CNRS, UMR7095, Institut d'Astrophysique de Paris, F-75014, Paris, France}
  

\begin{abstract}
We propose a new method to probe for variations in the fine structure constant $\alpha$ using clusters of galaxies, opening up a window on a new redshift range for such constraints. Hot clusters shine in the X-ray mainly due to bremsstrahlung, while they leave an imprint on the CMB frequency spectrum through the Sunyaev-Zel'dovich effect. These two physical processes can be characterized by the integrated Comptonization parameter $Y_{SZ}D_A^2$ and its X-ray counterpart, the $Y_X$ parameter. The ratio of these two quantities is expected to be constant from numerical simulations and current observations. We show that this fact can be exploited to constrain $\alpha$, as the ratio of the two parameters depends on the fine structure constant as $\propto \alpha^{3.5}$. We determine current constraints from a combination of Planck $SZ$ and XMM-Newton data, testing different models of variation of $\alpha$. When fitting for a constant value of $\alpha$, we find that current constraints are at the 1\% level, comparable with current CMB constraints. We discuss strategies for further improving these constraints by almost an order of magnitude.
\end{abstract}

\maketitle
\section*{Introduction}
\label{intro}
Constraints on the fine structure constant are currently derived from a number of different observations, ranging from laboratory to astrophysical measurements (see e.g.\cite{alphaconstraints}). E.g. CMB data from the WMAP7 satellite in combination with ACT and SPT data constrain $\alpha$ at $\sim 1\%$ level (68\% c.l.) at $z\sim 1000$ \cite{menegoni12} (see also \cite{scoccola12, landau10}). Other probes, such as the measurement of the CMB temperature  through the Sunyaev-Zel'dovich (SZ) effect \cite{sunyaev72} in clusters of galaxies \cite{avgoustidis12} (in a model dependent way) or 21cm absorption \citep{khatri07} can potentially probe variations of $\alpha$ at lower redshifts than CMB. Opening up new redshift ranges is useful as  theory is not a reliable guide to the expected nature of variations in fundamental constants, so that variations might be e.g. non-monotonic with $z$.
In this paper, we propose to constrain the fine structure constant combining measurements of the SZ effect with the measurement of X-ray emission in clusters of galaxies.

CMB experiments such as Planck \cite{planck}, SPT \cite{spt} or ACT \cite{act} are in fact currently detecting many hundreds of clusters through the SZ effect. Some of these are known clusters, while others are newly discovered, and have been or will soon be observed in follow-up campaigns by other observatories, such as  the Chandra or the XMM-Newton telescopes in the X-rays. Thus, measurements of  both the SZ effect  and of the X-ray emission of hundreds of clusters will soon be available up to a redshift of $z\sim 1$ \cite{chamballu10}.

The SZ effect is often expressed in terms of the integrated Compton parameter $Y_{SZ}D_A^2$ (see Sect. \ref{szeffect} for a detailed definition), while the X-ray emission of hot clusters ($kT\gtrsim 2KeV$), mainly due to bremsstrahlung, can be characterized by the parameter $Y_X=M_gT_X$ (see Sect. \ref{xrays}). Both $Y_{SZ}D_A^2$ and $Y_X$ are approximations of the thermal energy contained in the clusters, and are thus expected to strongly correlate with total mass, weakly depending on its dynamical state \cite{kaiser86, kravtsov06, borgani09}.
In the limit where gravity completely dominates cluster formation, $Y_{SZ}D_A^2$ and $Y_X$  are expected to scale in the same way with mass and redshift as power-laws.
Indeed, numerical simulations suggest they both have equivalent scaling relations that are close to be self-similar. Thus, the $Y_{SZ}D_A^2-Y_{X}$ relation is expected to be, on average, constant at all $z$ \cite{nagai06,kravtsov06,stanek10,fabjan11,kay11,bohringer12}. Furthermore, the same simulations show that the scatter on the relation between $Y_{SZ}D_A^2$ and $Y_X$ is small, at $\lesssim 15\%$ level.

So far, the data are consistent with these predictions. In particular, the Planck collaboration found no deviation from a self-similar behaviour of the $Y_{SZ}D_A^2-Y_X$ relation
using Planck SZ data and XMM-Netwon X-ray data for 62 clusters in the redshift range $0\lesssim z\lesssim 0.4$ \cite{planckeszselected}. 
Compatible results are also found  by \cite{rozoplanckchandra}, analyzing a subsample of 28 clusters from the Planck SZ catalog observed in the X-ray by the Chandra telescope. Additionally, the SPT collaboration confirmed these results using SZ data from the SPT telescope and Chandra X-ray data for 14 clusters \cite{spt}. 
These analyses do not find evolution of the relation with redshift.  \footnote{Rozo et al. \cite{rozoI} do find a hint of evolution with $z$, at the $3\sigma$ level, in the scaling relations found by the Planck collaboration but attribute this effect not to physics but to systematic effects in the analysis of the X-ray data.     }

In this paper, we propose to use the observed linear relation between $Y_X$ and $Y_{SZ}D_A^2$ to constrain the fine structure constant. In fact, the $Y_{SZ}$ and the $Y_X$ parameters have different 
dependencies on the fine structure constant, so that their ratio strongly depends on $\alpha$. The fact that no deviation from a constant have yet  been observed in the $Y_{SZ}D_A^2-Y_X$ relation can be used to constrain variations in $\alpha$.

Currently, the most powerful probe of $\alpha$ in the redshift range $0.3\lesssim z\lesssim 4$ are atomic absorption lines in quasar spectra.
Tantalizing hints of  variation of $\alpha$ have actually been found in this kind of data. A first  analysis of  143 absorption systems over redshift range $0.3 < z < 4.2$ observed by the Keck telescope suggested a time variation of the fine structure constant at the 5-sigma level \cite{murphy03}, a result that was however questioned by an independent analysis  (see e.g. \cite{srianand04}). A more recent result \cite{webb10, king12}, that combines the
Keck data with a new sample of 154 measurements from the VLT suggests on the other hand a spatial variation in $\alpha$. This spatial variation fits a dipole with declination $(-58\pm9)^\circ$ and right ascenscion $(17.5\pm0.9)h$. The significance of this claim is $4.2\sigma$, and no systematics that could justify such a variation have yet been found. 
This clearly stimulates the attempts to find additional hints of variation of $\alpha$, preferably with probes alternative to quasars. 
Clusters can explore redshifts lower than quasars, ($z\lesssim 1$ versus $ z\lesssim 4$), but these still correspond to a large range of look-back time ($ct(z=1)=7.7 Gly$ versus ($ct(z=1)=12.1 Gly$)). Furthermore, current quasar data provide tight contraints on the variation of the fine structure constant at redshifts higher than $\gtrsim 0.3$ \cite{webb10}, while clusters can already provide constraints at lower redshifts.  Potentially, clusters could be a better probe at low $z$, as there are many thousands of clusters at these redshifts, where it is harder to find quasar absorption systems.
 It is thus worth investigating  how much clusters can contribute in constraining $\alpha$.
 
We describe the dependence of the SZ effect on the fine structure constant in Sec. \ref{szeffect}, while the dependence of the X-ray data is described in Sec.~\ref{xrays}. The combination of the X-ray and SZ effect is described in \ref{szx}. In Sec.~\ref{currentdata} we derive constraints from a set of clusters observed, for the $SZ$ effect, by the Planck satellite and in the X-rays by the XMM-Newton telescope. We present constraints both in case $\alpha$ is assumed to be the same for the whole sample of clusters, and in case $\alpha$ is assumed to vary with space position and look-back time.
We conclude in Sec.~\ref{conclusions}.
\section{Method}

\subsection{Sunyaev-Zel'dovich effect}
\label{szeffect}
Over 80\% of the baryonic content of clusters of galaxies is expected to be under the form of intergalactic hot gas at temperatures of order $T\sim10^7-10^8 K$ \cite{borgani09}. 
The ionized gas  can inverse Compton scatter CMB photons, leaving a signature, the Sunyaev-Zel'dovich effect, in the CMB spectrum \cite{sunyaev72}. This spectral distortion is proportional to the Compton parameter $y$, that quantifies the gas pressure of the intracluster medium integrated along the line of sight:

\begin{eqnarray}
 y&=&\frac{\sigma_T}{m_ec^2}\int n_e(r) T(r)dl \label{yparam}\\&=&\frac{\sigma_T}{m_ec^2}\int_b^{R_B} n_e(r) T(r)\frac{r dr}{\sqrt{r^2-b^2}},\label{yparamII}
\end{eqnarray}
where $m_e$ is the electron mass, $n_e$ is the electron number density at distance $r$ from the center of the cluster, $T$ is the temperature of the gas, $b$ is the projected distance from the center of the cluster, $R_B$ is the radial extent of the cluster and $\sigma_T$ is the Thompson cross section, which depends on the fine structure constant as
\begin{equation}
 \sigma_T=\frac{8\pi}{3}\frac{\hbar^2}{m_e^2 c^2}\alpha^2.
\label{sigmat}
\end{equation}

Integrating the Compton parameter over the angular extent of the cluster  provides the integrated Compton cylindrical parameter $Y^{cyl}_{SZ}$,
\begin{eqnarray}
Y^{cyl}_{SZ}(R)D_A^2&=\int_0^R y(b)2\pi b db\sim D_A^2\int_0^{\theta(R)} y(\theta)2\pi \theta d\theta \nonumber\\
\end{eqnarray}
where  we expressed the projected distance as $b=\theta D_A$, with $\theta$ is the angular dimension of the cluster and $D_A$ the angular diametre distance, defined as
\begin{eqnarray}
D_A(z)&=&(1+z)^{-1}\int \frac{dz'}{H(z')}\\
\left[\frac{H(z)}{100 {\rm Km/s/Mpc}}\right]^2&=&\omega_m (1+z)^{3}+\omega_{\Lambda}\nonumber \\
\end{eqnarray}
Here, $\omega_m=\Omega_mh^2$ is the physical matter density and $\omega_\Lambda=\Omega_\Lambda h^2$ is the physical dark energy density. In the following , we will assume a flat $\Lambda CDM$ cosmology.

The cylindrical parameter can then be linked (see e.g. \cite{arnaud10}) to the commonly used spherical integrated Compton parameter, defined as:
\begin{eqnarray}
Y^{sp}_{SZ}(R)D_A^2&=&\frac{\sigma_T}{m_ec^2}\int_0^R n(r)T(r) 4\pi r^2dr\\\label{Ysz}
\end{eqnarray}
We will use the spherical integrated Compton parameter in the following, calling it simply $Y_{SZ}$.
 
Thus the  $Y_{SZ}$ parameter depends on the fine structure constant via the Thompson cross section in Eq. \ref{sigmat} as
\begin{equation}
Y_{SZ}\propto \alpha^2                                                                                                                                                                                                                                                                                                                                                                                                   \end{equation}
\subsection{X-rays}
\label{xrays}
At the high temperatures of galaxy clusters, the intergalactic gas emits mainly through thermal bremsstrahlung, whose emissivity, i.e. the emitted power per unit volume at frequency $\nu$, is (see e.g. \cite{sasaki96,karzas61,sarazin88}):
\begin{eqnarray}
\epsilon_\nu&=&\frac{2^5\pi e^6}{3m_ec^3}\left(\frac{2\pi}{3m_ek}\right)^{1/2} Z^2 g_{ff} n_e n_i T^{-1/2} e^{-h\nu/kT}\nonumber\\
&=&\alpha^3 \frac{2^5\pi\hbar^3}{3m_e}\left(\frac{2\pi}{3m_ek}\right)^{1/2} Z^2g_{ff}  n_e n_i T^{-1/2} e^{-h\nu/kT}\nonumber\\
\label{bremis}
\end{eqnarray}
where $n_i$ is the ion density, $k$ the Bolzmann constant, $h$ the Planck constant, $Z$ is the atomic number. $g_{ff}(Z,T,\nu)$ is the Gaunt factor which corrects for quantum mechanical effects and varies slowly with frequency and temperature.
 
We do not observe  the emissivity directly, but the surface brightness, i.e. its integral along the line of sight at different angular distances from the center of the cluster $\theta$:
\begin{equation}
	I_\nu(\theta)=\int_{b}^{R_B} \frac{\epsilon_\nu(r)2 r dr}{\sqrt{r^2-b^2}}.
	%\frac{dL}{d\Omega dA}
\label{surb}
\end{equation}

Thus, deprojecting the surface brightness allows measuring the emissivity, which depends on the estimate of the angular diameter distance as $\epsilon_\nu \propto D_A^{-1}$.

The magnitude of the X-ray emission is often quantified through the  $Y_X$ parameter \cite{kravtsov06,arnaud10}, which is analogous to the SZ parameter $Y_{SZ}$. It is defined as
\begin{equation}
 Y_X=M_g(R) T_X \label{Yx}
\end{equation}

Here, $M_g(R)$ is the X-ray determined gas mass within a certain cluster radius $R$, and $T_X$ is the spectroscopically determined X-ray temperature of the cluster, determined within a cylindrical annulus. 
The gas mass is defined as
\begin{eqnarray}
M_g(R)&=&\mu_e m_p\int{n_e dV}\\&\propto& n_e R^3 \label{mgne}
\end{eqnarray}

with $\mu$ is the mean molecular weight of eletrons. The gas mass can be determined from X-ray data as the density profile of the cluster can be inferred from the emissivity in Eq. \ref{bremis}, which is in turn obtained from the observed surface brightness in Eq. \ref{surb}. This assumes that the temperature can be spectroscopically determined and that the angular diameter distance is known.

We can then link the inferred gas mass to the fine structure constant  from Eq. \ref{bremis} and \ref{mgne}
\begin{eqnarray}
M_g(R)&\propto&\sqrt{\epsilon_\nu T^{1/2}e^{h\nu/kT}\alpha^{-3}}R^{3}\\ &\propto&\sqrt{ \alpha^{-3}I_\nu D_A^{-1}} D_A^{3}\\
&\propto& \sqrt{  \alpha^{-3}I_\nu}D_A^{5/2}
\end{eqnarray}
  The dependence on the angular diameter distance derives from the dependence of $R=\theta D_A$ and from the fact that the emissivity is derived from the observed surface brightness, $\epsilon_\nu\propto D_A^{-1}$.

 The $Y_X$ parameter thus depends on the fine structure constant as
\begin{equation}
Y_X\propto \alpha^{-1.5}
\end{equation}

\subsection{$Y_{SZ}-Y_{X}$ relation and $\alpha$}
\label{szx}
From Eq. \ref{Yx}, \ref{Ysz} and \ref{mgne}, the ratio between  $Y_{SZ}$ and $Y_{X}$ depends on the structure of a cluster as
\begin{eqnarray}
\frac{Y_{SZ}D_A^2}{Y_X}&=&C_{XSZ}\frac{\int{n_e(r) T(r) dV}}{T_X(R)\int{n_e(r) dV}}\\
C_{XSZ}&=&\frac{\sigma_T}{m_e c^2}\frac{1}{\mu_e m_p} 
\end{eqnarray}

The $Y_{SZ}$ parameter depends on the gas mass weighted temperature, while $Y_X$ depends on the X-ray temperature. Both are approximations of the same physical quantity, i.e. the  thermal energy of the cluster.
It is then clear that if clusters were isothermal, the ratio between the two would exactely be equal to a constant. However, the ratio can still expected to be constant if the temperature profile of the clusters is universal. This condition is fullfilled if the evolution of clusters is completely dominated by gravity, weakly depending on gas physics. In this case, both $Y_{SZ}$ and $Y_X$ are expected to strongly correlate with the mass of the cluster via the virial theorem, both with the same dependence on mass and redshift \cite{kaiser86,borgani09}
 \footnote{This can be easily shown as follows. Assuming spherical simmetry, the mass  $M_{\Delta_c}$ included in a radius $r_{\Delta_c}$ within which the mean density is $\Delta_c$ times the critical density at redshift z,  $\rho_c(z)=\rho_c(0)(H_(z)/H_0)^2$, is: $$M_{\Delta_c}\propto\rho_c(z)\Delta_cr_{\Delta_c}^3.$$ For the virial theorem, $kT\propto \phi\propto M_{\Delta_c}/r_{\Delta_c}$, with $\phi$ the gravitational potential of the cluster. Thus $Y_{SZ},Y_X\sim M_{\Delta_c}T\propto M_{\Delta_c}^{5/3}(H(z)/H_0)^{2/3}$.}.
 
Numerical simulations  have shown that indeed the two parameters have scaling relations with the total mass of the cluster that are very close to be self-similar, i.e. that
$Y_X, Y_{SZ} \propto M^{5/3}E(z)^{2/3}$, with $E(z)=(H(z)/H_0)$. Furthermore, they have shown that the scaling relation between $Y_X$ and $Y_{SZ}D_A^2$ has very small scatter, at the level of $\sim 15\%$\cite{stanek10,kay11,fabjan11}. The relation between the two is also expected to be independent of redshift, as their scaling relation with mass have the same dependence on cosmology.
Finally, the relation seems not to crucially depend on the dynamical state of the clusters \cite{arnaud10}.

Based on all these considerations, the ratio between $Y_{SZ}$ and $Y_{X}$ is expected to be constant
 $$\frac{Y_{SZ}D_A^2}{Y_X}\sim const$$  
for clusters at different redshifts or space positions.

This fact can be exploited to constrain the variation of the fine structure constant at different time/space positions $i$, as
\begin{equation}
 \left(\frac{Y_{SZ}D_A^2}{Y_X}\right)_i=\left(\frac{\alpha_i}{\alpha_0}\right)^{3.5}\left(\frac{Y_{SZ}D_A^2}{Y_X}\right)_0
\end{equation}
where $\left(\frac{Y_{SZ}D_A^2}{Y_X}\right)_0$ is a reference value of the ratio that assumes a reference value of the fine structure constant $\alpha_0$. The method enables us to measure the relative variation of $\alpha$ with respect to $\alpha_0$ as a function of redshift and space position. 
Alternatevely, if one could reliably estimate a reference value of $\left(\frac{Y_{SZ}D_A^2}{Y_X}\right)_0$ knowing the value of $\alpha_0$, e.g. from simulations, it would also be possible to have an absolute measure of $\alpha$ for each cluster.

In any case, if a variation is detected, it could be clearly either due to an uncorrected astrophysical or instrumental systematic error  or due to an actual change in $\alpha$. But if no variation is detected a limit on the variation of $\alpha$ can be extracted. We cannot logically exclude the  possibility that intrisic changes of the ratio $\left(\frac{Y_{SZ}D_A^2}{Y_X}\right)$ or uncorrected systematics  might provoke a variation in the $Y_{SZ}-Y_X$ relation that conspires to precisely cancel a true  variation in $\alpha$ resulting in no apparent variation.This case would lead to a false rejection of the variation hypothesis.

\section{Constraints from current data}
\label{currentdata}
\subsection{Data}
We present in this section constraints on $\alpha$ from current data.
For the analysis, we use SZ and X-ray data from a subsample of the Planck Early Sunyaev-Zel'dovich cluster
sample \citep{planckesz}, as reported in \cite{planckeszselected}. The clusters of the ESZ sample are detected in the Planck all-sky maps through their thermal SZ imprint on the CMB. They are characterized by a S/N higher than $6$, and are required to have a X-ray counterpart in the MCXC catalog \cite{piffaretti11}. The subsample then reported in \cite{planckeszselected} is composed by 62 clusters that had been observed by the XMM-Newton telescope, that are not contamined by flares and whose morphology is regular enough that spherical symmetry can be assumed. We additionally exclude  from the analysis cluster A2034, whose redshift estimate is discordant in \cite{planckeszselected} and \cite{mantz10}, as noted by \cite{rozoI}. We thus use 61 clusters, in the redshift range $0.044<z<0.44$. The $Y_{SZ}$ and $Y_X$ parameters we use here are measured within a radius $R_{500}$, i.e. the radius at which the mean matter density of the cluster is $500$ times larger than the critical density  at the redshift of the cluster. Furthermore, X-ray temperatures are defined within a cylindrical annulus of radius $[0.15-0.75]R_{500}$.

\begin{figure*}[ht!]
\centering
\includegraphics[angle=0,width=0.45\textwidth]{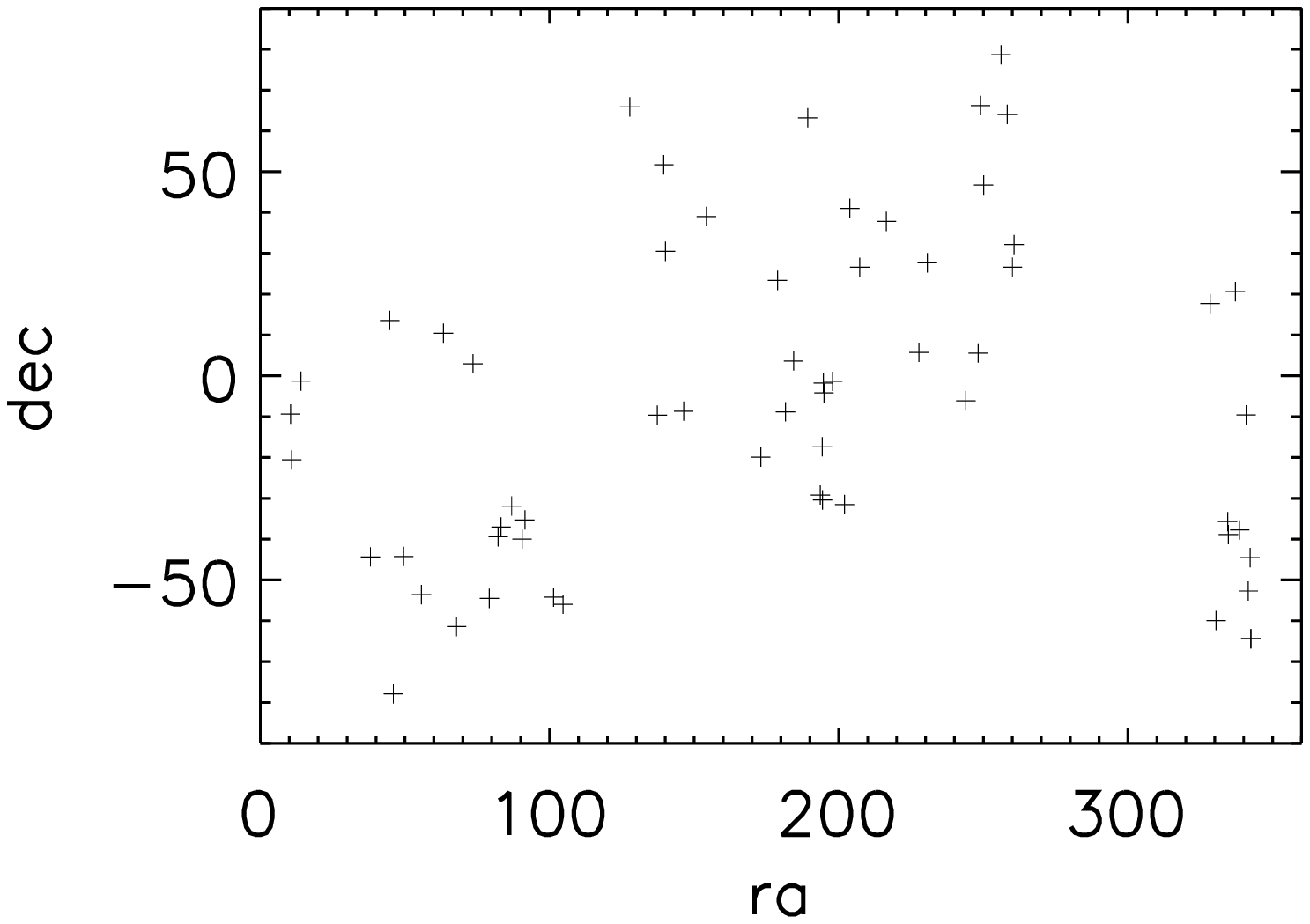}
 \includegraphics[angle=0,width=0.45\textwidth]{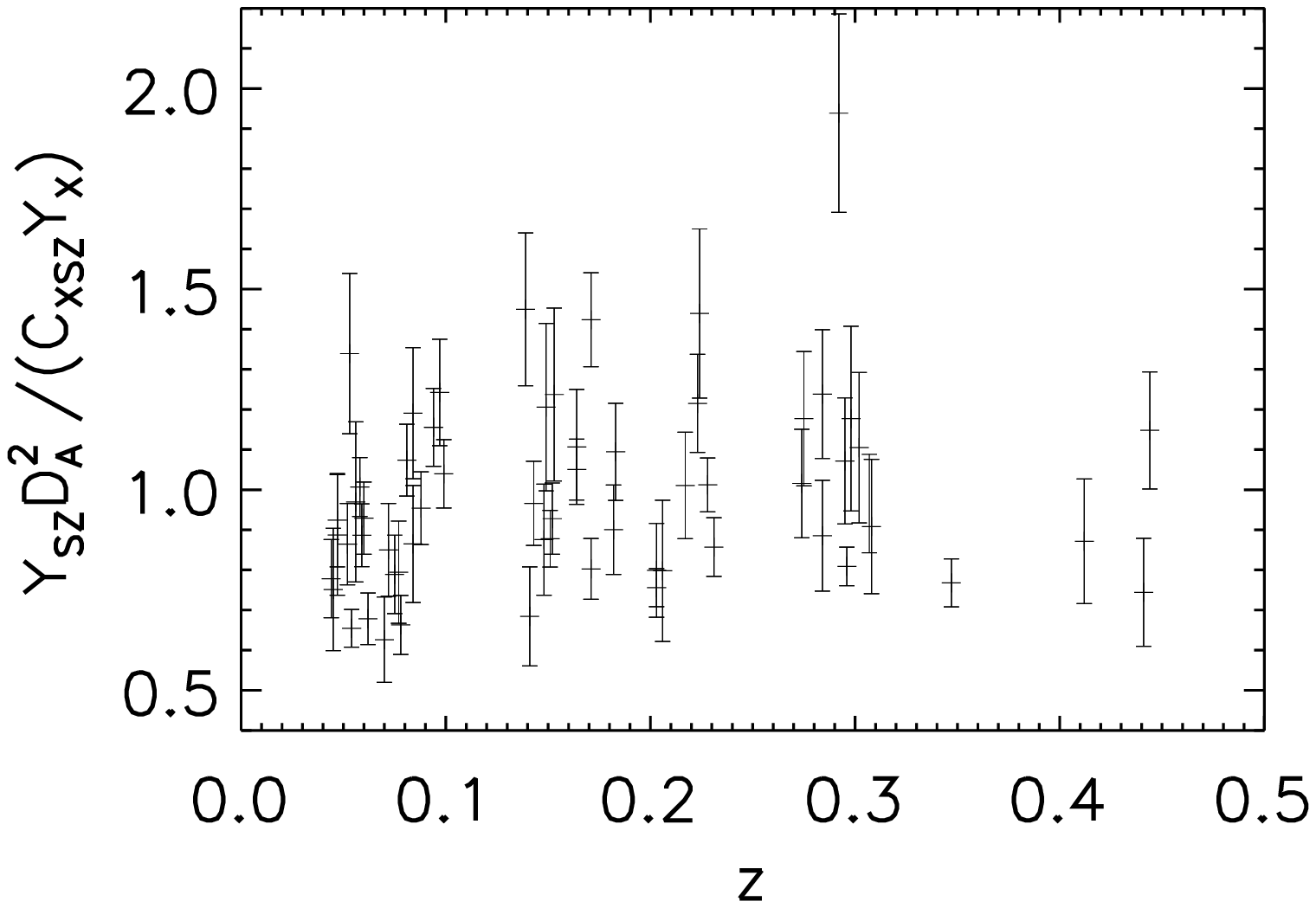}
\caption{Left: Right ascenscion and declination of the Planck ESZ cluster subsample used for the analysis. Right: $Y_{SZ} D_A^2/C_{XSZ}Y_X$ of the clusters in function of redshift. The error bars shown are calculated from error propagation of the errors on $Y_{SZ}$ and $Y_{X}$ as published in \cite{planckeszselected}.}
\label{clustersvsz}
\end{figure*}

Fig. \ref{clustersvsz} shows the space and redshift distribution of the clusters used.

These data are neither a complete nor a representative sample of clusters, and the observation of a larger sample of clusters in the X-ray will be required to properly characterize the Planck clusters, in particular to study the intrisic scatter and Malmquist bias, as well as possible systematics \cite{planckval3}. However, we use this dataset to provide a first estimate of the constraints on $\alpha$ that one can derive from this dataset. 
\subsection{Analysis and constraints: Constant $\alpha$}
We  first analyze the data in order to find constraints on $\alpha$ under the simple assumption that no evolution in time or space is present, i.e. that the $Y_{SZ}D_A^2/Y_X$ ratio is a constant. Any deviation exceeding statistical error is attributed to intrinsic scatter.

We calculate the mean of $Y_{SZ}D_A^2/Y_X$ through a modified weighted least square method (MWLS). This method differs from a simple weighted least square because it takes into account the fact that statistical uncertainties on $Y_{SZ}D_A^2/Y_X $, calculated by propagating the statistical errors on $Y_{SZ}D_A^2$ and $Y_X$, can be underestimated or can neglect intrinsic scatter. A weighted least square method provides in fact a simple weighted mean of $\log{(Y_{SZ}D_A/C_{XSZ}Y_X)}=-0.056\pm 0.01$, with a $\chi^2$ per degrees of freedom of $\chi^2/dof=233/60$. Clearly, such a high reduced $\chi^2$  might indicate either that a constant is a poor description of the $Y_{SZ}D_A^2/Y_X$ data, or the presence of e.g. additional intrinsic scatter. Under this second assumption, we can account for a possible wider dispersion of the data by quadratically adding to the statistical error of each data point a constant term $\sigma_{\rm intr}$(see e.g. \cite{pratt06}) for the unknown intrinsic scatter. The weighted mean and the intrinsic scatter are then jointly determined so that the reduced $\chi^2$ equals  $1$. 
Following this method, we obtain $\log{(Y_{SZ}D_A/C_{XSZ}Y_X)_i}=-0.056\pm 0.03$, i.e.,$(Y_{SZ}D_A/C_{XSZ}Y_X)_i=0.96\pm 0.027$. This result is in perfect agreement with the results found by \cite{planckeszselected} and \cite{rozoI}. 	We find that the intrinsic  scatter term for each data point is equal to $\sigma_{\rm intr}=0.18$. We note here that we 
do not correct the data for Malmquist bias, which for this set of data is not expected to modify the best fit \cite{planckeszselected}, but might provide a slightly higher estimate of the intrinsic scatter compared to corrected data.

In order to check the results from the MWLS method, we also calculate the mean as a simple  arithmetic average and estimate its uncertainty by bootstrap resampling. In this case we obtain $(Y_{SZ}D_A/C_{XSZ}Y_X)=0.96\pm 0.021$, in agreement with what previously found. The scatter in this case is calculated following \cite{planckeszselected}: we calculate the scattering term $\sigma_{\rm intr}'$ as the quadratic difference between the raw scatter $\sigma_{raw}$ and the statistical uncertainty
 \begin{eqnarray}
\chi^2_r&=&\sum_i \frac{(x_i-<x_i>)^2}{\sigma^2(x_i)}\frac{1}{dof} \\ 
	\sigma^2_{stat}&=&\frac{1}{N}\sum_i \sigma^2(x_i)\\	
 \sigma_{raw}^2&=&\chi^2_r  \frac{N}{\sum_i 1/\sigma^2(x_i)} \\
 	\sigma_i'^2&=&\sigma_{raw}^2-\sigma^2_{stat}
 	\end{eqnarray}
 where $\chi^2_r$ is the reduced $\chi^2$, $dof$ is the number of degrees of freedom, in this case equal to $dof=60$, and $N$ is the number of clusters, in this case equal to $N=61$.  The recovered scatter is $\sigma_{intr}'=0.17\pm 0.026$, in perfect agreement with what found with the first method. The uncertainty on the scatter is calculated as in \cite{planckstatistical}, $\Delta(\sigma_{\rm intr}')=\sigma^2_{\rm intr}(2N(N-1))^{-1}\sum(1+\sigma(x_i)/\sigma_{\rm intr^2})^2.$
   
  We can then convert these results to a measurement on $\alpha$. The current assumption is that $(Y_{SZ}D_A^2/C_{XSZ}Y_X)$ ratio is constant, and thus that the fine structure constant has the same value for all the clusters considered, $\alpha=\alpha_0$. The uncertainty  on $\alpha$ is then: $$\frac{\sigma(\alpha)}{\alpha_0}=\frac{1}{3.5}\frac{\sigma(Y_{SZ}D_A^2/C_{XSZ}Y_X)}{(Y_{SZ}D_A^2/C_{XSZ}Y_X)_0}=0.0077,$$ i.e. a constraint on $\alpha$ at $\sim 0.8\%$ level at 68\% c.l. This uncertainty includes both statistical error and  intrinsic scatter, but does not include uncertainties on the cosmological parameters used to determine the angular diameter distance, here chosen for our reference cosmology, i.~e.~flat $\Lambda CDM$  with  $H_0=70 \rm Km/s/Mpc$, $\Omega_m=0.3$, and $\Omega_\Lambda=0.7$.

The cosmological parameters are not perfectly known, and degeneracies with $\alpha$ could limit the constraining power of clusters. We now analyze the impact of these uncertainties.

The dependence of the $(Y_{SZ}D_A^2/C_{XSZ}Y_X)$ ratio on the angular diameter distance is
\begin{eqnarray}
\left(\frac{Y_{SZ}D_A^2}{C_{XSZ}Y_X}\right)_{ref}&\sim &\left(\frac{Y_{SZ}D_A^2}{C_{XSZ}Y_X}\right)_{true} \left(\frac{(D_A)_{ref}}{(D_A)_{true}}\right)^{-0.5}
\end{eqnarray}
where $ref$ indicates the angular diameter distance calculated with the reference cosmology, and $true$ indicates the unknown true cosmology.

First, a wrong estimate of the angular diameter distance could generate a "fake" evolution with redshift of the $(Y_{SZ}D_A^2/C_{XSZ}Y_X)$ ratio.
Second, the uncertainties on the knowledge of $D_A$ can affect the errors on $\alpha$. Still, the dependence is weak and current constraints on the angular diameter distance are  at the level of a few percent \cite{dawson12,larson11}. We thus expect that, at least for current data, the uncertainty on cosmological parameters should not affect constraints on $\alpha$.

In order to quantify this statement, we redetermine the uncertainty on $\alpha$ through the Fisher Matrix methodology, marginalizing over cosmological parameters (see e.g. \cite{tegmark}).
Together with the fine structure constant, we consider as parameters the physical matter density $\Omega_mh^2$ and the Hubble constant $H_0$. We impose  priors on
$\Omega_mh^2$ and on the Hubble constant $H_0$ from a combination of CMB data (WMAP7, ACBAR, SPT, ACT),  the galaxy power spectrum extracted from the SDSS-DR7 luminous red galaxy sample and the constraints on $H_0$ from the Hubble Space Telescope observations, obtained when a $\Lambda CDM+\alpha$ model is fitted to the data \cite{menegoni12}. The priors are $\sigma^P(\Omega_mh^2)=0.021$ and $\sigma^P(H_0)=2.1 \rm Km/s/Mpc$. 
The Fisher matrix for ${Y=Y_{SZ}{D_A^2}/Y_X}$ is
\begin{equation}
 F^{Y}_{ij}=\sum_{k=1}^{N} \frac{\partial Y_k }{\partial\theta_i} \frac{1}{\sigma(Y_k)^2} \frac{\partial Y_k} {\partial\theta_j}
\label{fisherY}
\end{equation}
where $\theta$ are the parameters and $\sigma(Y_k)$ is the error on the $k$ data point, that includes the statistical uncertainty and the intrinsic scatter on $Y_k$ previously determined. To calculate the derivatives, we assume a fiducial true cosmology equal to the reference one, with $\alpha=\alpha_{ref}=\alpha_0$.

Adding the priors, the total Fisher Matrix is
\begin{equation}
F_{ij}=F_{ij}^Y+P_{ij} \quad \text{with}\quad P_{ij}=\frac{1}{(\sigma^p_{ii})^2}\delta_{ij}
\end{equation}
where $\sigma^p_{ii}$ is the prior on parameter $ii$.

The resulting error on $\alpha$ from the Fisher Matrix is given by $(F^{-1/2})_{\alpha\alpha}$, so that $\sigma(\alpha)/\alpha_0=0.0086$, while we recover the previous result when assuming the cosmological parameters perfectly known.
Thus we conclude that for current data, the uncertainties on the cosmological parameters only marginally affect the constraint on $\alpha$, at the level of $\sim 10\%$.  

As more and more clusters are found, this might become a limiting  factor for constraints on $\alpha$ from clusters. 
We estimate that  cosmological parameter uncertainties become the dominant source of uncertainty on $\alpha$ once $\sim 6000$ clusters are observed. In this case the constraint on  $\alpha$ is $\sigma(\alpha)/\alpha_0=0.003$. On the other hand, upcoming data from on-going experiments such as Planck are expected to improve the constraints on cosmological parameters. In particular \cite{martinelli12} forecasts the constraints on cosmological parameters from a combination of future CMB and weak lensing data for a Planck satellite-like experiment and a Euclid satellite-like experiment, for a model where  $\alpha$ is also allowed to vary. In this case the forecast uncertainties for the Hubble parameter and the matter density are $\sigma^p(H_0)=0.34$ and $\sigma^p(\omega_m)=0.0007$. With $6000$  clusters the constraint on $\alpha$ would improve to $\sigma(\alpha)/\alpha_0=0.001$. 

\subsection{Testing the dipole}
\begin{figure}[t!]
\includegraphics[angle=0,width=0.45\textwidth]{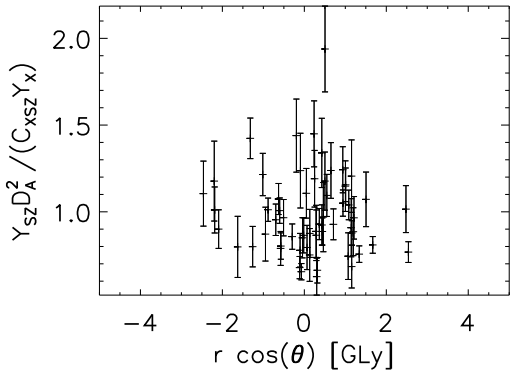}
\caption{$Y_{SZ}D_A^2/C_{XSZ}Y_X$ in function of $r \cos(\theta)$ for the 61  clusters used in the analysis. }
\label{clustersvsrcotheta}
\end{figure}

We now present results for a fit of the data that allows for a variation of $\alpha$ in space and with lookback-time. In particular, we consider the same model of variation used by  \cite{webb10} to fit the quasar data. This is useful in order to compare the constraining power of clusters to quasars, as the two sets of data have different redshift and space distributions. The model of variation we adopt for $\alpha$ is
\begin{equation}
\frac{\alpha}{\alpha_0}=Ar\cos(\theta)+1
\end{equation}
where $r$ is the lookback time, $$r=\int\frac{dz'}{H(z')},$$ $\theta$ is the angular distance from the best fit dipole and $A$ is the amplitude of the effect. Thus, the relation we fit is of the type
\begin{eqnarray}
\label{fittingmodel}
\left(Y\right)^{1/3.5}&=& (a+b rcos\theta)\nonumber\\
a&=&Y_0^{1/3.5}\nonumber\\
b&=&Y_0^{1/3.5}A
\end{eqnarray}
We fix here the position of the dipole to the best fit found by \cite{webb10}, so that we only fit for $a$ and $b$ with the MWLS method, again allowing for intrinsic scatter. The results on the derived parameters of interest are $Y_0=0.96\pm 0.028$ and $A=(-5.5\pm 7.9)\times 10^{-3}GyL^{-1}$. The constraint on $A$ is much weaker than the one obtained from quasars, namely $A=(1.1\pm0.25)\times10^{-6}GyL^{-1}$.

As in the previous section, we verify that this constraint is not currently affected by uncertainties on cosmological parameters. With the same Fisher Matrix procedure previously described, marginalizing over the uncertainties on $\Omega_m h^2$ and $H_0$, we obtain $\sigma(Y_0)=0.031$ and $\sigma(A)=8.1\times10^{-3}$. Thus, also in this case uncertainties on cosmological parameters impact the constraints at the $\sim 10\%$ level.
These findings indicate that current cluster data cannot directly probe the quasar claim, at least under the assumption that the variation is an increasing function of redshift as in Eq. \ref{fittingmodel}. This is however a phenomenological assumption that is not justified by any theoretical model. Thus, variations of $\alpha$ could have a  more complicated evolution with time. Clusters of galaxies could potentially unveil variations at low redshifts.

\section{Conclusions}
\label{conclusions}
We propose a new method to constrain the fine structure constant by using SZ and X-ray observations, opening a complementary redshift window on $\alpha$.
With $61$ clusters in our data set, no evolution has been detected in the scaling relation between the integrated Compton parameter $Y_{SZ}D_A^2$ and the X-ray analogous parameter $Y_X$ so far.  We can take advantage of this fact to put a constraint on the fine structure constant in the redshift range $0\lesssim z\lesssim 0.5$ at the $0.8\%$ level. The ratio between the two parameters have in fact a strong dependence on the fine structure constant, namely $(Y_{SZ}D_A^2/Y_X)\propto \alpha^{3.5}$. 

This constraint is not limited by degeneracies with cosmological parameters at this point. It is conservative since we assume that the intrinsic X-SZ relation does not evolve. If there were a systematic evolution it would be absorbed in our estimate of the intrinsic scatter which softens the constraint. 

Further improvements to this constraint are possible. Cosmological parameter degeneracies are subdominant until $6000$ clusters are observed, at which point the constraint would improve to $0.3\%$. The intrinsic scatter in the $Y_{SZ}/Y_{X}$ relation could potentially be reduced with improved modeling of the physics in the cluster gas, leading to a further improvement. This is currently the dominant factor limiting the constraints.
 Also clusters are unlikely to improve significantly the constraints on spatial variation of $\alpha$. 
Ultimately a combination of increased numbers of clusters with improved cosmological parameters constraints from upcoming large surveys such as Planck and Euclid could sharpen the constraint to $\lesssim 0.1\%$. This forecast takes cosmological parameter degeneracies fully into account.
\acknowledgments

We would like to thank for very fruitful  comments and discussions Benjamin Wandelt, Marta Volonteri, Carlos Martins, Alessandro Melchiorri, Nabila Aghanim and James Bartlett.
This work was supported by Benjamin D. Wandelt's ANR Chaire d'Excellence ANR-10-CEXC-004-01.

\end{document}